\documentclass[%
 amsmath,amssymb,
prx,
twocolumn
]{revtex4-2}

\usepackage{graphicx}
\usepackage{dcolumn}
\usepackage{bm}

\usepackage{microtype}
\usepackage{booktabs} 
\usepackage{comment}
\usepackage[dvipsnames]{xcolor}
\definecolor{myblue}{RGB}{35, 123, 148}
\definecolor{myred}{RGB}{242, 145, 0}
\usepackage{hyperref}
\hypersetup{
	colorlinks=true,
	linkcolor=myred,
	filecolor=magenta,      
	urlcolor=myblue,citecolor=myblue}
\usepackage{braket}

\usepackage{amsmath}
\usepackage{amssymb}
\usepackage{mathtools}
\usepackage{amsthm}

\usepackage[capitalize]{cleveref}

\begin{document}


\title{Interpretable representation learning of quantum data enabled by probabilistic variational autoencoders}

\author{Paulin de Schoulepnikoff}
\affiliation{University of Innsbruck, Department for Theoretical Physics, Technikerstr. 21a, A-6020 Innsbruck, Austria}
\author{Gorka Mu\~noz-Gil} 
\email{gorka.munoz-gil@uibk.ac.at}
\affiliation{University of Innsbruck, Department for Theoretical Physics, Technikerstr. 21a, A-6020 Innsbruck, Austria}
\author{Hendrik Poulsen Nautrup} 
\affiliation{University of Innsbruck, Department for Theoretical Physics, Technikerstr. 21a, A-6020 Innsbruck, Austria}
\author{Hans J. Briegel} 
\affiliation{University of Innsbruck, Department for Theoretical Physics, Technikerstr. 21a, A-6020 Innsbruck, Austria}

\date{\today}

\begin{abstract}
Interpretable machine learning is rapidly becoming a crucial tool for scientific discovery. Among existing approaches, variational autoencoders (VAEs) have shown promise in extracting the hidden physical features of some input data, with no supervision nor prior knowledge of the system at study. Yet, the ability of VAEs to create meaningful, interpretable representations relies on their accurate approximation of the underlying probability distribution of their input. When dealing with quantum data, VAEs must hence account for its intrinsic randomness and complex correlations. While VAEs have been previously applied to quantum data, they have often neglected its probabilistic nature, hindering the extraction of meaningful physical descriptors. Here, we demonstrate that two key modifications enable VAEs to learn physically meaningful latent representations: a decoder capable of faithfully reproducing the probability distributions of quantum states and a probabilistic loss tailored to this task. Using benchmark quantum spin models, we identify regimes where standard methods fail while the representations learned by our approach remain meaningful and interpretable. Applied to experimental data from Rydberg atom arrays, the model autonomously uncovers the phase structure without access to prior labels, Hamiltonian details, or knowledge of relevant order parameters, highlighting its potential as an unsupervised and interpretable tool for the study of quantum systems.
\end{abstract}

\maketitle


\section{Introduction} \label{sec:intro}

The rapid progress of classical machine learning (ML) is profoundly impacting our ability to analyze quantum systems and process quantum data, paving the way for new paradigms in scientific discovery~\cite{iten2020dicvovering}. In the quantum domain, by leveraging both supervised and unsupervised learning methods, researchers are now able to identify quantum phase transitions~\cite{carrasquilla2017machine, cybinski2024speak, miles2023machine, huembeli2018identifying}, model complex entanglement features~\cite{linda2024predicting}, and improve quantum state tomography~\cite{torlai2018neural}. Generative ML in particular, has become an impactful tool in quantum physics, from the use of generative adversarial networks (GANs) for Hamiltonian learning ~\cite{kim2024hamiltonian} to diffusion models for quantum circuit synthesis \cite{furrutter2024quantum}. However, such models are typically black boxes, hindering our ability to interpret the origin or reasons of their predictions.

In this sense, ML techniques with enhanced interpretability can serve as a lens through which the intrinsic properties of quantum systems can be unraveled~\cite{wetzel2025interpretable}. For example, support vector machines have been employed to identify order parameters~\cite{ponte2017kernel}, symbolic regression methods have revealed underlying equations of motion~\cite{kharkov2021discovering}, and neural architectures leveraging correlation functions have been used to map out phase diagrams of experimental quantum many-body systems~\cite{miles2021correlator}. Among these interpretable models, variational autoencoders (VAEs)~\cite{kingma2013vae} have attracted attention due to their ability to autonomously extract meaningful and minimal representations of some input dataset in their \emph{latent space}~\cite{higgins2017betavae}. When applied to physics datasets, this relates to extracting the meaningful physical features of the system of study~\cite{iten2020dicvovering}. In the quantum realm, VAEs have been used for instance to study phase transitions in spin systems~\cite{walker2020deep,naravane2023semi}.

Nonetheless, one of the key challenges when applying machine learning methods to quantum data lies in its intrinsically stochastic nature. As a result, physical observables must be inferred statistically over repeated measurements.
This is particularly challenging in experimental platforms such as ultra-cold atoms in optical lattices~\cite{ebadi2021quantum} or trapped-ion quantum simulators~\cite{li2023probing}, where performing repetitive measurement is costly, often leading to poor statistics.
This fundamental stochasticity has motivated the development of ML approaches that operate directly on individual measurement outcomes, bypassing the need for full state reconstruction and instead learning meaningful features from single or ensembles of quantum snapshots~\cite{patel2022unsupervised, miles2023machine, cybinski2024speak, suresh2025interpretable}.

In this sense, accounting for the intrinsic stochasticity of quantum data is essential for developing interpretable machine learning models. Since the physical properties of a quantum system are fundamentally encoded in its probability distribution, capturing this distribution is necessary for any meaningful characterization. This is particularly critical for VAEs:  if they cannot reproduce the properties of the underlying probabilistic distribution of its input, it will not be able to create meaningful representations in its latent space.
However, existing VAE applications on quantum data have not considered this stochasticity~\cite{walker2020deep,wetzel2017unsupervised, jang2024unsupervised, naravane2023semi, frohnert2024explainable}. In particular, they commonly use deterministic losses, which fundamentally misaligns with the goal of learning the probability distribution of the target quantum states. Such losses direct the model to capture only one-body, or \textit{mean-field}, features, thereby neglecting nontrivial many-body correlations.
It is hence critical to develop architectures that can cope with the stochastic properties of the quantum data, if one aims to use this method to extract physical features. 

To tackle these challenges, we propose two key adaptions to the VAE architecture. First, we train the model to directly approximate the underlying probability distribution of the input data, rather than exactly reconstructing particular samples. Second, we adapt the VAE's decoder to be able to capture the stochastic properties of effectively any target quantum state. To achieve the latter, we draw inspiration from Neural Quantum States (NQS)~\cite{carleo2017solving}, neural network architectures used as expressive variational ansätze capable of representing quantum probability distributions (see Ref.~\cite{lange2024from} for an in depth review on the topic).

To demonstrate the impact of the proposed improvements into the VAE framework for quantum data, we compare its resulting performance to that of a standard deterministic VAE on two paradigmatic spin models: the transverse-field Ising model with next-nearest neighbor interactions (NNN-TFIM) and long-range interactions (LR-TFIM). Through these models, we will explore and identify regimes where conventional VAEs struggle while the presented method provides physically interpretable representations. We further validate the approach on experimental snapshots of Rydberg atom arrays~\cite{ebadi2021quantum}. Without access to any prior labeling, knowledge of the governing Hamiltonian nor the number of relevant order parameters, the model is able to autonomously uncover the underlying phase structure, underscoring its potential as a powerful unsupervised and interpretable tool for the study of quantum many-body systems.

\section{Methods} \label{sec:methods}



In this section, we provide a detailed description of the proposed pipeline, schematically presented in~\cref{fig:proposed_archi}. We consider here a quantum system that can be tuned through two controllable parameters $\theta_1$, $\theta_2$ (see \cref{fig:proposed_archi}a). A measurement device records snapshots for a wide range of parameter values. This dataset of snapshots, which contains no information about the parameters $\theta_{1,2}$, is then used to train a VAE. As we will explain below, the VAE learns a compact, physically meaningful representation of the system's physical descriptors. This representation can be accessed by inspecting the value of the latent neurons $\textbf{z}$ for snapshots across the parameter space (see \cref{fig:proposed_archi}b). As we will show in ~\cref{sec:results}, such latent representations will directly correlate with known order parameters of the systems under study, allowing us to identify different phases or regions throughout the parameter space.
In the following, we outline the fundamental principles of VAEs, as well as the key adaptations required to effectively handle the stochastic nature of quantum data.

\subsection{Background on Variational Autoencoders} \label{sec:VAE_background}

Variational Autoencoders (VAEs) consist of three main components: an encoder, a latent space, and a decoder~\cite{kingma2013vae}. The encoder compresses the input data into a lower-dimensional representation, namely the latent space. In contrast to autoencoders, the latent space of VAEs is parameterized by a probability distribution, typically defined as a multivariate Gaussian distribution. In practice, the encoder outputs the mean $\mu_i$ and variance $\sigma_i$ of the latent variables from which the values for each latent neuron $z_i$ are sampled and subsequently passed to the decoder (see \cref{fig:proposed_archi}c). The decoder then reconstructs the input data based on the sampled latent variables. 

The objective of Variational Autoencoders is to approximate an unknown probability distribution $p(x)$, assumed to underlie a given dataset. Directly computing this likelihood is generally intractable for high-dimensional data, so VAEs optimize a tractable lower bound known as the Evidence Lower Bound (ELBO). From the previous, the loss with which the model is then trained can be written as
\begin{equation} \label{eq:loss_VAE}
    \mathcal{L} = \frac{1}{N}\sum_{n=1}^{N}\mathbb{E}_q[\log p(x^{(n)}|z)] - \beta \mathrm{KL}[q(z|x^{(n)})||p(z)]
\end{equation}
with $N$ being the number of samples $x^{(n)}$ in the training set.

The loss function introduced in \cref{eq:loss_VAE} contains two terms. First, a reconstruction loss, which enforces similarity between the input and output data. Second, a regularization term, which minimizes the Kullback-Leibler (KL) divergence between the latent space distribution $q(z_i)\sim \mathcal{N}(\mu_i,\sigma_i)$ and the prior $p(z)$, typically defined as standard normal distribution $p(z)\sim \mathcal{N}(0,1)$. 
This term acts as a regularizer, enforcing the latent representation to align with the prior distribution. However, if all latent variables were to collapse to the prior, the decoder would lack sufficient information to faithfully reconstruct the data, leading to an increase in the reconstruction loss. 

This trade-off induces a natural competition between the two terms in the loss function, such that only the minimal set of latent neurons needed to properly reconstruct the input diverge from the prior (commonly referred to as an \emph{active} neuron), while the other neurons are "noised out" and collapse to $p(z)$ (referred to as \emph{passive} neurons). Moreover, one can further scale the regularization term with a hyperparameter $\beta$~\cite{higgins2017betavae}, such that the two terms are properly balanced.

In practice, this mechanism forces the model to retain only the essential information from the input data in the latent space. This means that the VAE will noise out unnecessary latent neurons until finding a minimal set able to describe such input data. When the latter is associated with a few hidden variables, e.g. the degrees of freedom of a physical system, the resulting latent representation tends to align with these~\cite{iten2020dicvovering, nautrup2022operationally, fernandez2024learning}.

\subsection{VAE with a probabilistic decoder}

\begin{figure*}[t]
    \centering
    \includegraphics[width=0.8\textwidth]{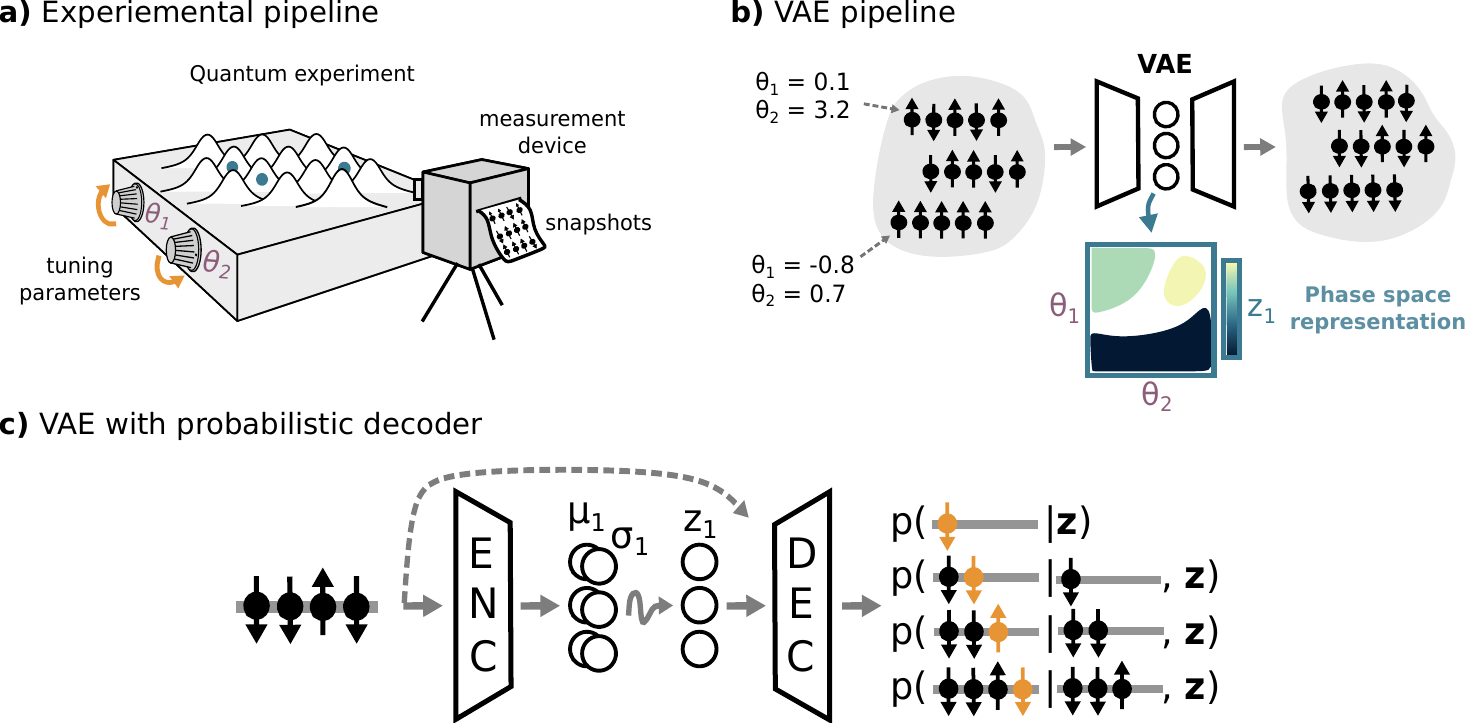}
    \caption{\textbf{Schematic representation of the proposed pipeline.} 
    \textbf{a)} An experimental setup produces snapshots of a quantum system for different values of the experimental parameters $\theta_{1,2}$. \textbf{b)} A variational autoencoder (VAE) is trained on an ensemble of unlabeled snapshots collected from the previous experiment across a wide parameter range. Inspecting the learn latent neurons $\mathbf{z}$ across the parameter space will shed light about the system's phase space (see \cref{sec:results} for more). 
    \textbf{c)} Schematic representation of the cpVAE. The encoder neural network, ENC, takes as input a spin configuration. The latter outputs the means $\mu_i$ and the variances $\sigma_i$  parameterizing the Gaussian distribution from which the latent variables $z_i$ are sampled. Then, the decoder neural network DEC takes as input the latent variables and outputs the conditional probabilities of each spin be in a given state (up or down in the present scheme). During training, the decoder also receives as input the spin configuration to be reconstructed by the conditional probabilities, as indicated with the dotted arrow. Once trained, the latent neurons $z_i$ encode the main physical features of the physical system, which directly relate to its phase space. Moreover, one can fix the value of the latent variables and generate new configurations by autoregressively sampling the output conditional probabilities.}
    \label{fig:proposed_archi}
\end{figure*}

As discussed in the introduction, prior applications of VAEs to quantum data have largely employed architectures with deterministic decoders~\cite{walker2020deep,wetzel2017unsupervised, jang2024unsupervised, naravane2023semi, frohnert2024explainable}. In these models, the decoder is trained to reconstruct individual spin configurations using a deterministic reconstruction loss between input and output spin configurations, as for instance the mean squared error (MSE). This approach implicitly assumes a deterministic relationship between the latent representation and the output configuration, neglecting the intrinsic probabilistic nature of quantum measurement outcomes. Moreover, the use of a deterministic reconstruction metric enforces the model to learn average configurations across samples rather than capturing the full probability distribution of outcomes. This leads to a latent space that reflects only the mean-field structure of the data, failing to capture any probabilistic feature, as for instance correlations between nearby spins. In what follows, we refer to this baseline model as the \emph{deterministic} VAE (dVAE) to distinguish it from the stochastic decoder architecture introduced in this work.

Motivated by the recent work on VAEs for stochastic processes~\cite{fernandez2024learning} and arguments on using generative classifiers for phase-classification~\cite{arnold2024mapping}, we present here the key changes that will allow VAE to correctly create interpretable representations of quantum data. First, we construct a decoder inspired by the recent advances in Neural Quantum States (NQS) \cite{lange2024from}. For example, NQS have recently demonstrated impressive performance in ground-state searches and non-equilibrium dynamics~\cite{chen2025convolutional}. By integrating an NQS-based decoder into the VAE framework, we ensure that the model can accurately capture and regenerate the statistical structure of the input quantum states. 

Second, we propose replacing the deterministic loss with a probabilistic reconstruction loss. Rather than attempting to reproduce individual measurement outcomes, which is fundamentally intractable due to the inherent stochasticity of quantum systems, we instead train the model to learn the underlying probability distribution from which these input samples are drawn. Indeed, when dealing with a quantum system, a single measurement outcome cannot be used to fully characterize it. Instead, one needs to model the full distribution over all possible outcomes. By aligning the decoder's objective with this principle, the model becomes capable of capturing the essential statistical structure of the quantum state, rather than averaging over its fluctuations.

To account for the two previous arguments, we propose the use of an auto-regressive neural network (ARNN) for the decoder (see \cref{fig:proposed_archi}c and \cref{an:NN_archi} for further details), and train it to model the conditional probabilities of each spin configuration, given the preceding spins.  More precisely, the configuration of spin $i$ is given by the Bernoulli distribution. Its probability mass function $f$ over the two possible outcomes $x_i\in \{1,-1\}$ is 
\begin{equation}
    f(p_i,x_i) = p_i^{x_i-1}(1-p_i)^{x_i+1},
\end{equation}
with $p_i$ being predicted by the decoder. Without loss of generality, the whole spin configuration is written as the product of the conditional probabilities of each spin,
\begin{equation}
    p(x) = \prod_{i=1}^N p(x_i|x_0, x_1, ..., x_{i-1}).
\end{equation}
Hence, each output of the decoder models a conditional probability
\begin{equation}
    p_i := p(x_i|x_0, x_1, ..., x_{i-1}). 
\end{equation}
We will refer to this model as cpVAE, standing for \emph{conditional probabilistic} VAE. Given these considerations, the reconstruction loss (first term in \cref{eq:loss_VAE}) can be rewritten as
\begin{align}
    \mathbb{E}_{z\sim q}[\log p(x_n|z)] = & \mathbb{E}_{z\sim q} \sum_i (x_i-1)\log p_i \nonumber \\
    & + (x_i+1)\log (1-p_i).
\end{align}

To compute the KL regularization (second term in \cref{eq:loss_VAE}), we adopt the decomposition introduced in the Total Correlation Variational Autoencoder (TC-VAE) framework~\cite{chen2018isolating}. This approach enables a fine control over the structure of the latent space, by explicitly penalizing statistical dependencies between latent variables. This enforces the distinct degrees of freedom of some input physical data to be stored in independent latent neurons. The ELBO TC-decomposition consists in rewriting the KL regularization as
\begin{align}
    \mathbb{E}_{p(x)}\big[ \mathrm{KL(q(z|x)||p(z))} \big] = & \mathrm{KL}(q(z,x)||q(z)q(x)) \nonumber \\
     & + \mathrm{KL}(q(z)||\prod_jq(z_j)) \label{eq:TC}\\
     & + \sum_j \mathrm{KL}(q(z_j)||p(z_j)). \nonumber 
\end{align}
In the latter, the first term on the right-hand side is the mutual information $I_q(z;x)$. The second term, referred to as the total correlation (TC), encourages the model to learn as so-called disentangled representations~\cite{chen2018isolating}, where individual latent units capture distinct, minimally correlated physical features of the data. The third term is the dimension-wise KL which aligns the latent variables with the prior $\mathcal{N}(0,1)$. The hyperparameters, $\alpha$, $\beta$ and $\gamma$ are introduced  to control the weight of each of these terms. The full loss with which the model is trained is thus
\begin{align} 
    \mathcal{L} = & \mathbb{E}_{z\sim q} \sum_i (x_i-1)\log p_i + (x_i+1)\log (1-p_i) \nonumber\\
    & - \alpha I_q(z;x)  - \beta \mathrm{KL}(q(z)||\prod_jq(z_j)) \label{eq:final_loss} \\
    & - \gamma \sum_j \mathrm{KL}(q(z_j)||p(z_j)). \nonumber
\end{align}

\section{Results} \label{sec:results}

To assess the performance of the proposed cpVAE architecture, we begin by benchmarking its performance against a dVAE on two well-established quantum spin models. These controlled settings will allow us to identify the limitations of deterministic decoders, as for instance their inability to capture non-trivial features beyond the mean-field behavior. They will also demonstrate that the proposed probabilistic approach yields representations that more accurately reflect the underlying physics.

The first spin model is the transverse-field Ising model (TFIM) with next-nearest-neighbor (NNN) interaction. For weak NNN coupling, the model effectively reduces to the conventional TFIM, a setting where prior studies have applied deterministic VAEs~\cite{walker2020deep, wetzel2017unsupervised, jang2024unsupervised}. In this regime,  we show that this success is largely due to the fact that the relevant order parameter is a mean-field observable. As the strength of the NNN interaction increases, the system deviates from such mean-field behavior, and we observe that the deterministic VAE fails to reconstruct physical features, in contrast to the probabilistic one, effectively hindering the representation of the system in their latent space.

We further compare both models on the TFIM with long-range interactions (LR-TFIM), where the interplay between quantum fluctuations and non-local couplings poses an additional challenge. In this setting, we show that a probabilistic variant of VAE is essential for capturing the relevant physical features in the latent space.

Finally, we apply our architecture to experimental data obtained from Rydberg atom arrays. Despite the complexity and stochastic nature of the measurements, we will show that the model autonomously uncovers meaningful latent representations that reflect the phase structure of the system, without any supervision or prior knowledge of the underlying Hamiltonian.

The code needed to reproduce the results presented in each figure is available in the accompanying code repository~\cite{github}.

\subsection{Spin Models} \label{sec:results_spins}

\subsubsection{NNN-TFIM}

\begin{figure}
    \centering
    \includegraphics[width=0.9\columnwidth]{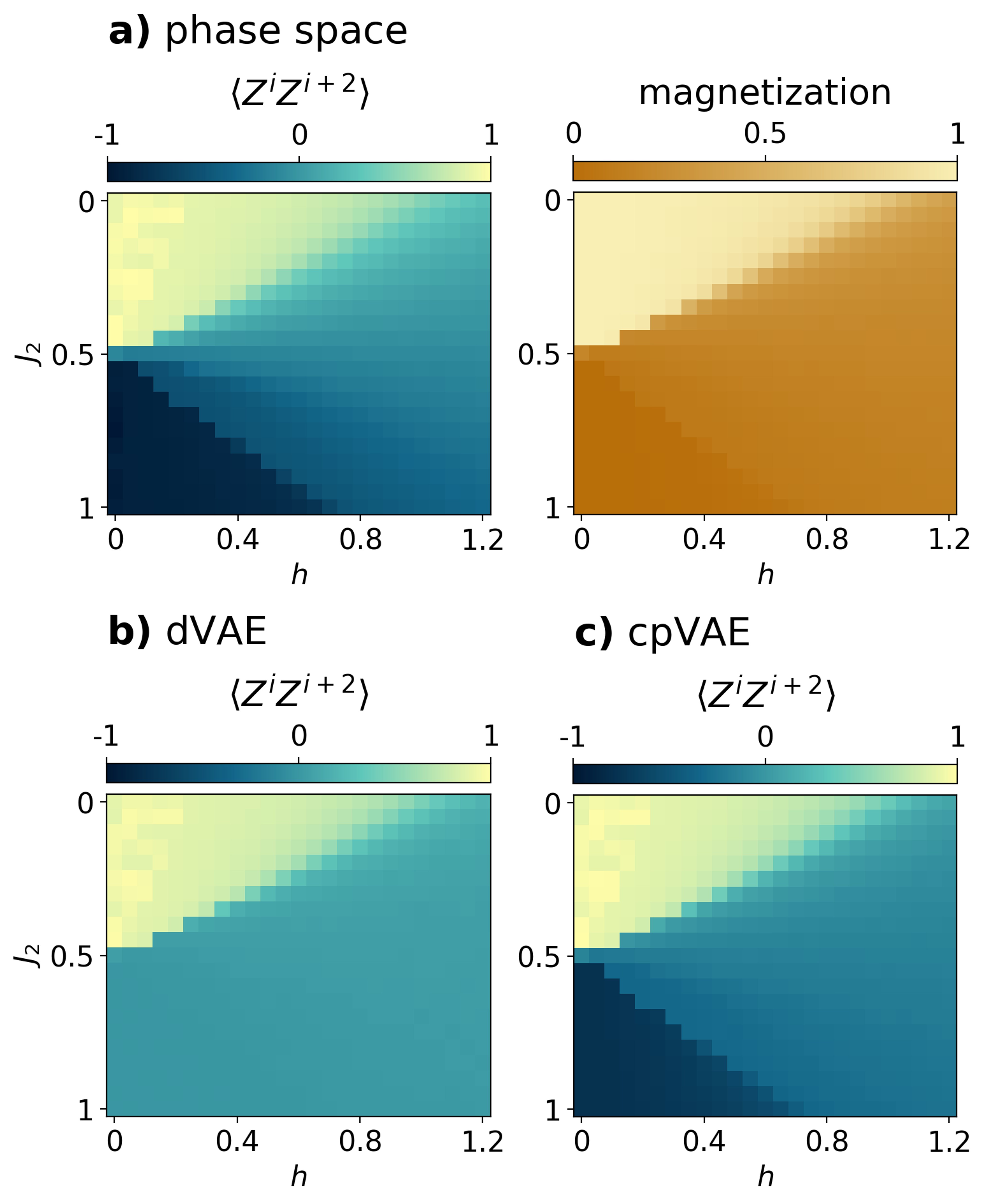}
    \caption{\textbf{Next-nearest-neighbor transverse-field Ising model.}
    \textbf{a)} Next-nearest-neighbor correlator (left) and  magnetization (right) computed from training set configurations across the phase space.  
    \textbf{(b,c)} Next-nearest-neighbor correlator of the generated spin configurations across the phase space generated by the dVAE and cpVAE, respectively.}
    \label{fig:NNNTFIM_phase_space_and_corr2_reconstr}
\end{figure}

We start by considering the Transverse Field Ising Model with nearest neighbor and next-nearest-neighbor interaction (NNN-TFIM) in a one dimensional chain, whose Hamiltonian has the form\cite{kassan2001one}
\begin{equation} \label{eq:H_NNNTFIM}
    H_{NNN-TFIM} = \sum_{i=1}^N -Z^iZ^{i+1} + J_2Z^iZ^{i+2} + h X^i,
\end{equation}
with $J_2 \in [0,1]$ and $X^i, Z^i$ being the Pauli operators on spin $i$. The system under consideration has a size of $N = 20$ with periodic boundary conditions.

The phase diagram of the NNN-TFIM can be characterized using two observables: the next-nearest-neighbor correlator $\langle Z^iZ^{i+2}\rangle$ and the magnetization $\langle Z^i \rangle $, as shown in \cref{fig:NNNTFIM_phase_space_and_corr2_reconstr}a.
For coupling strengths $J_2<0.5$, the nearest-neighbor interaction dominates, and the system effectively reduces to the conventional TFIM, a regime where deterministic VAEs had previous success~\cite{walker2020deep, wetzel2017unsupervised, jang2024unsupervised}. 

As $J_2$ increases, the next-nearest-neighbor antiferromagnetic interaction becomes dominant. Due to the one-dimensional topology of the lattice, this term induces a geometric frustration. In particular, for $J_2>0.5$ and small transverse field $h$, the system enters an antiferromagnetic phase characterized by alternating strong and weak nearest-neighbor correlations. 

We then train the cpVAE with configurations arising from different points in the phase space of this spin model. During training, we see that a single neuron remains active, while all other neurons collapse to the prior and are noised out (see \cref{an:training} for further details). This is to be expected since, as shown in \cref{fig:NNNTFIM_phase_space_and_corr2_reconstr}a, the whole phase space can be characterized only by the second nearest-neighbor correlator. To mitigate the effect of hyperparameter tuning and to make the comparison as fair as possible, we then train the dVAE without KL regularization and with a latent dimension of 1.  By doing so, we substantially ease its training, and ensure that its latent representation is minimal. We note that not doing so, and considering a larger latent space, would allow the dVAE to store some stochastic features of the input sample in the latent space, for example, particular spins being up or down. While the KL term may help to find a more interpretable representation, our aim here is not to compare the structure of the latent representations, but rather to demonstrate that the dVAE cannot learn parts of the phase space due to its deterministic nature, in contrast to the cpVAE. A more detailed explanation of this benchmark strategy can be found in~\cref{an:training}.

\paragraph*{Reconstruction analysis} 
We then benchmark the generative capabilities of both models. To do so, we first input configurations from the whole phase space to the encoder, extract the corresponding latent space, and then sample new configurations from the decoder based on the extracted latent variables. In \cref{fig:NNNTFIM_phase_space_and_corr2_reconstr}b, we show the $\langle Z^iZ^{i+2}\rangle$ correlator for the output configurations of the dVAE across the phase. 

While the model accurately reproduces the expected correlations in both the ferromagnetic and paramagnetic regimes, it fails to capture the antiferromagnetic phase. Importantly, this regime cannot be characterized by the magnetization only, the usual order parameter extracted in previous works. Since the alternating bond structure inherent to the frustrated ground state falls beyond any mean-field parameter, the deterministic decoder is unable to correctly reproduce this statistics, and hence the antiferromagnetic phase will never appear in the latent representation of a dVAE.

In contrast, we show in \cref{fig:NNNTFIM_phase_space_and_corr2_reconstr}c that the cpVAE is able to accurately reproduce the desired statistics across the full phase space, including the antiferromagnetic region. Notably, this accuracy is maintained even when the model is trained on a restricted subset of the parameter space (see \cref{an:NNNTFIM_generalize}), demonstrating its capacity to generalize beyond the training domain. This highlights the ability of the cpVAE to capture many-body correlations that elude deterministic approaches, reinforcing the importance of a probabilistic generative model when learning from intrinsically stochastic quantum data.

\subsubsection{LR-TFIM}

\begin{figure*}
    \centering
    \includegraphics[width=\linewidth]{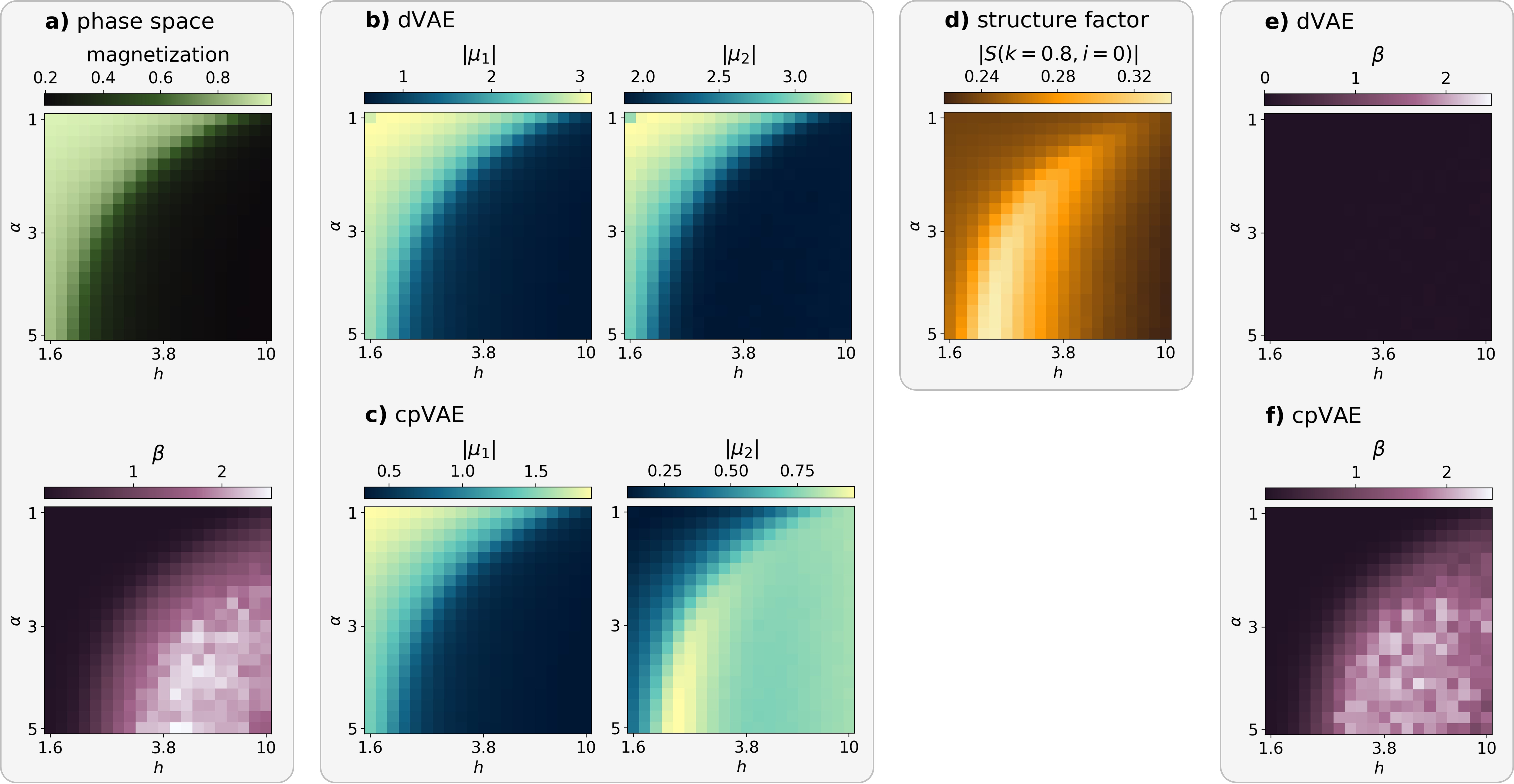}
    \caption{\textbf{LR-TFIM}. \textbf{a)} Magnetization (top) and $\beta$ exponent from \cref{eq:corr_exp} (bottom) computed from training set configurations across the phase space.  
    \textbf{(b,c)} Absolute value of the learned latent space mean values $\mu_i$ for the two active neurons, for the dVAE and cpVAE, respectively, and for input configurations across the phase space. 
    \textbf{d)} Structure factor from \cref{eq:struc_factor} with $k=0.8$ and $i=0$ for training set configurations across the phase space. \textbf{(e,f)} $\beta$ exponent for spin configurations generated by the dVAE and the cpVAE, respectively. The former has values $\beta=0$ for across the whole phase space.}
    \label{fig:LRTFIM_phase_space_latent_reconstr}
\end{figure*}

As a second benchmark, we consider the long-range transverse field Ising model (LR-TFIM), which has been recently performed experimentally via trapped-ion simulators~\cite{hauke2013spread, jurcevic2014quasiparticle, li2023probing}. This model is a modification of the conventional TFIM in which spin–spin interactions decay algebraically with distance. Its Hamiltonian is given by
\begin{equation} \label{eq:H_LRTFIM}
    H_{LR-TFIM} = \sum_{i=1}^N hX^i - \sum_{j\neq i}\frac{Z^iZ^j}{|i-j|^\alpha},
\end{equation}
where $\alpha$ controls the range of the interactions: smaller values of $\alpha$ correspond to a slower decay and hence longer-range couplings. To better reflect typical experimental conditions, we simulate the system using open boundary conditions. A chain of $N=20$ spins is considered.

At small transverse field strengths $h$, the ground state is ferromagnetic, with all spins aligned in the same direction. For large $h$, the spins are aligned in the direction of the transverse field, leading to a paramagnetic phase. The ground state becomes then a product state devoid of long-range correlations. These two phases can be clearly visually differentiated by means of the total magnetization, as we show in~\cref{fig:LRTFIM_phase_space_latent_reconstr}a. Moreover, for intermediate values, the system exhibits quantum fluctuations with nontrivial many-body entanglement and algebraically decaying spin–spin correlations. The latter can be expressed as
\begin{equation}
\label{eq:corr_exp}
    C(r) = \langle Z^iZ^{i+r} \rangle \sim \frac{1}{r^{\beta(\alpha,h)}},
\end{equation}
where the exponent $\beta$ is an indicator of the correlation strength and spread.

Training the cpVAE on configurations arising from this spin model, we observe that two latent neurons survive (see~\cref{an:training}). As in the previous section, the dVAE was trained without KL regularization and imposing the same latent dimension found with the cpVAE. 

\paragraph*{Latent space analysis}
In this case, rather than focusing on the output of the models, we will directly analyze their latent space, to better understand the representations learned by the two variants. To do so, we input configurations sampled at different points of the phase space and analyze the mean value of the two surviving neurons $\mu_i$ output by the encoder. 

The latent space learned by the deterministic VAE (dVAE) is shown in \cref{fig:LRTFIM_phase_space_latent_reconstr}b. Due to the inability of the decoder to properly account for the correlations of the model, the latent neurons of dVAE follow the same behavior as the total magnetization. Indeed, both neurons exhibit similar behavior, effectively collapsing onto a representation that distinguishes only the ferromagnetic phase. As a result, the dVAE fails to capture any meaningful structure beyond mean-field characteristics, such as those arising in the critical or intermediate regimes.

In contrast, the latent space learned by the cpVAE (\cref{fig:LRTFIM_phase_space_latent_reconstr}c) exhibits a markedly richer structure. The first latent neuron, $\mu_1$, correlates strongly with a magnetization-like observable (see \cref{an:LRTFIM_sampling} for more details), which is consistent with it capturing the dominant factor of variation in the data. The second, $\mu_2$, encodes a distinct feature that highlights the presence of an intermediate phase emerging at low values of $\alpha$ and $h$. To investigate this further, we applied symbolic regression~\cite{crammer2023interpretable} to the latent values of $\mu_2$, seeking a compact analytical expression $f(x)$ that best describes its behavior given the corresponding spin configuration $x$  (see~\cref{an:SR_LRTFIM} for more details). This approach suggests that the neuron was encoding a quantity similar to the structure factor~\cite{baez2020dynamical}
\begin{equation}
\label{eq:struc_factor}
    S(k,i) = \frac{1}{N}\sum_j e^{2\pi k|j-i|/N}\langle Z^jZ^i \rangle .
\end{equation}
As shown in \cref{fig:LRTFIM_phase_space_latent_reconstr}d, the structure factor, evaluated here at fixed momentum $k = 0.8$ and spatial reference site $i = 0$, highlights the same region of the phase diagram as captured by the latent variable $\mu_2$.  From a physical perspective, the structure factor reveals a distinct intermediate regime between the ferromagnetic phase and the high $\beta$ paramagnetic regime. This feature, visible only at the chain boundaries (i.e., $i=0$ and $i=N$), suggests the emergence of a boundary-induced phase, likely arising from finite-size and open-boundary effects~\cite{vodola2015long}. That the cpVAE identifies this behaviour without supervision underscores its potential as a tool for uncovering nontrivial structure in many-body systems and guiding the formulation of new hypotheses.

\paragraph*{Reconstruction analysis}
To further demonstrate the importance of the probabilistic decoder, we now perform the same analysis on generated configurations as done in the previous section. We then compute the correlation exponent $\beta$ from \cref{eq:corr_exp}. As expected, the dVAE fails to reproduce the spatial correlations present in the data, yielding spin configurations with $\beta=0$ across the entire phase space, hence without the expected polynomial decay. In contrast, the cpVAE correctly recovers the spatial structure of the quantum states. The reconstructed configurations exhibit $\beta$ exponents in quantitative agreement with the ground-state correlations across the entire phase space. This result confirms that the cpVAE not only correctly captures mean-field features such as magnetization but also encodes nontrivial many-body correlations.

\subsection{Experimental Rydberg Atoms} \label{sec:results_Rybderg}

As a final benchmark, we train the cpVAE with experimental snapshots obtained from a programmable Rydberg atom array~\cite{ebadi2021quantum}.  The setup consists of a two-dimensional square lattice with $N=L\times L$ atoms, where each atom resides in either its electronic ground state $\ket{g}$ or an excited Rydberg state $\ket{r}$. The latter is achieved via laser-driven excitation of an outer electron to a high-lying energy level, giving rise to strong, long-range interactions between excited atoms. This platform provides an ideal testbed for the cpVAE, as it exhibits a variety of quantum phenomena, including frustration, quantum criticality and emergent spatial order, all encoded in the projective measurement snapshots. Our aim here is to assess whether cpVAE can extract meaningful latent representations from such data, despite the presence of experimental noise and the stochastic nature of quantum measurement.

\begin{figure*}[t]
    \centering
    \includegraphics[width=1.\linewidth]{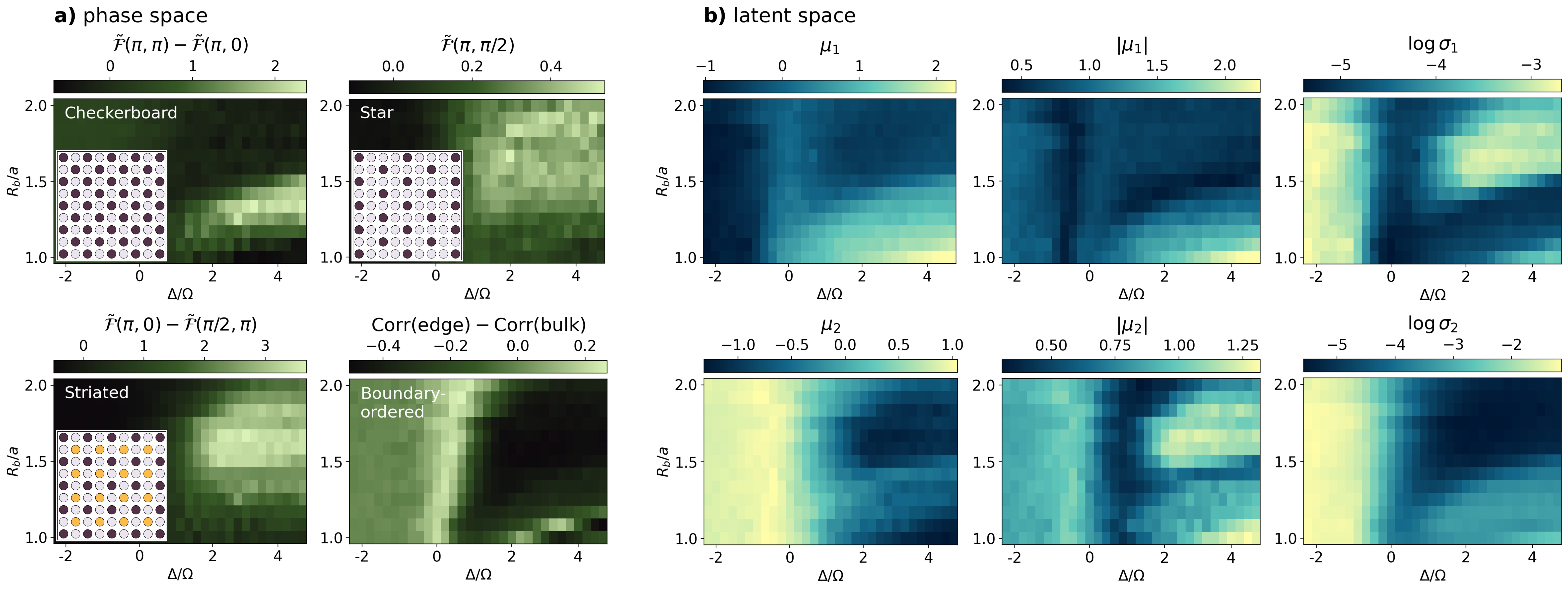}
    \caption{\textbf{Experimental Rydberg atoms array.} 
    \textbf{a)} Different Fourier-space order parameters (\cref{eq:fourier}) highlighting the location of the checkerboard, star and striated phases across the phase space. The boundary ordered phase is located by means of the difference between the nearest-neighbor correlator for atoms at the edge and the bulk of the array.
    The insets showcase exemplary configurations of each phase on a smaller $9\times 9$ lattice, with white, purple and orange circles representing atoms in the $\ket{g}$, $\ket{r}$ and $\ket{+_x}$ states, respectively.  
    \textbf{b)} Latent representation learned by the cpVAE. For the two remaining active neurons, we present here the mean values (left) and variances (right) as output by the encoder, for input configurations across the phase space. We also show the absolute mean values (center), which showcase a finer resolution of the learned latent space.}
    \label{fig:Rydberg_phase_space_latent_space}
\end{figure*}

The Rydberg array under study is characterized by the Hamiltonian

\begin{align}
    H_{\mathrm{Rydberg}} = & \sum_{i<j} \frac{C_6}{||\mathbf{r_i}-\mathbf{r_j}||^6} \hat{n}_i\hat{n}_j - \Delta\sum_i \hat{n}_i \nonumber \\
    & - \frac{\Omega}{2} \sum_i \sigma_i^x
\end{align}
with
\begin{equation}
    \hat{n}_i = \frac{1}{2}(\sigma_i^z+\mathcal{I}) \hspace{0.5cm} \mathrm{and} \hspace{0.5cm} C_6 = \Omega \Big(\frac{R_b}{a}\Big)^6,
\end{equation}
where $\sigma_i^x= \ket{g}_i\bra{r}_i + \ket{r}_i\bra{g}_i$, $\sigma_i^z= \ket{r}_i\bra{r}_i - \ket{g}_i\bra{g}_i$, $\Delta$ is the detuning from resonance, $\Omega$ the Rabi frequency, $a$ the lattice length scale and $R_b$ the Rydberg blockade radius. A key feature of Rydberg atoms is that, when a given atom is in the excited state $\ket{r}$, the van der Waals potential effectively prevents the excitation of any other atom within a radius $R_b$, an effect commonly referred to as the "Rydberg blockade"~\cite{lukin2001dipole,urban2009observation}. The experimental setup considered here, consisting of an array of $13\times13$ atoms, allows for the tuning of two different parameters: $R_b/a$, controlling the strength of the blockade, and $\Delta/\Omega$ encoding the relative strength of a longitudinal field. 
To create the snapshots, the neutral atoms went though a coherent quantum evolution for different values of blockade strength $R_b/a$ and relative field strength $\Delta/\Omega$. 
Then, a projective readout is performed in order to determined which atoms were not in the excited Rydberg state $\ket{r}$~\cite{miles2023machine, ebadi2021quantum}.

The phase space of the described system is illustrated in \cref{fig:Rydberg_phase_space_latent_space}a. First, three different phases, namely the checkerboard, star and striated phases can be identified through the different Fourier-space order parameters~\cite{ebadi2021quantum}
\begin{equation}
    \tilde{\mathcal{F}}(k_1,k_2) = \langle \mathcal{F}(k_1,k_2) +\mathcal{F}(k_2,k_1) \rangle /2  
\end{equation}
with 
\begin{equation}
\label{eq:fourier}
     \mathcal{F}(\mathbf{k}) = \Bigg|\frac{1}{\sqrt{N}}\sum_i^{N_{data}} e^{i\mathbf{k}\cdot\mathbf{x}_i/a}n_i\Bigg|.
\end{equation}
Each panel in \cref{fig:Rydberg_phase_space_latent_space}a shows the particular parameter used to identify each phase (see panels' titles), with insets showcasing exemplary configurations of each phase on a smaller $9\times 9$ lattice. Moreover, an edge-ordered phase can be also recognized by the difference between the nearest neighbor correlator of atoms at the edge and at the bulk of the array. For $R_b/a > 1.5$ and between the edge-ordered and the star phase, an additional ordered phase was identified in Ref.~\cite{miles2023machine}.

We train the cpVAE on experimental snapshots collected across a range of control parameters $R_b/a$ and $\Delta/\Omega$ (see details in \cref{an:training}). We observe for this dataset that only two latent neurons remain active. Their behaviour across the phase diagram is presented in \cref{fig:Rydberg_phase_space_latent_space}b, where we plot their mean $\mu_i$, absolute means $|\mu_i|$ and variances $\sigma_i$ (see \cref{fig:proposed_archi}). 

The left columns, corresponding to $\mu_i$, reveal that the latent dimensions are sufficient to discriminate between distinct quantum phases. For example, the regime in which the system energetically favors minimizing the number of atoms in the $\ket{r}$ state ($\Delta/\Omega < 0$) is characterized by low $\mu_1$ and high $\mu_2$ values. Then, the star phase shows low values of $\mu_1$, the striated phase low values in both components, and the checkerboard phase exhibits large $\mu_1$.  By analyzing the absolute values $|\mu_i|$ (central columns of \cref{fig:Rydberg_phase_space_latent_space}b), we found an even better resolved picture of the latent space, where the boundary-ordered phase is clearly highlighted for $|\mu_1| = 0$. 

The right columns of \cref{fig:Rydberg_phase_space_latent_space}b display the variances $\sigma_i$ of the latent dimensions, which quantify the uncertainty associated with each neuron's encoding. Lower values of $\sigma_i$ correspond to narrower posterior distributions, implying reduced variability in the sampled latent variable $z_i$ around its mean $\mu_i$. While the KL regularization term in the loss function encourages larger $\sigma_i$ to match the unit-variance prior $p(z)$, this pressure is counteracted by the reconstruction loss, which favours more precise sampling. In regions of the phase diagram where redundant or non-discriminative information is encoded, the model can tolerate increased variance without compromising reconstruction accuracy. This effect is visible, for example, at $\Delta/\Omega<0$, corresponding to nearly no atoms in the $\ket{r}$ state, where $\sigma_i$ becomes large. By inspecting the variance across the latent dimensions, one gains insight into their relative importance in different regions of the latent space. In particular, the first latent shows reduced variance in the vicinity of the boundary-ordered and checkerboard phases, indicating its relevance in encoding these features. Conversely, the second latent exhibits small variance around the star and striated phases, suggesting its role in resolving these configurations. A similar study of the variances of the cpVAE trained on the spin model is provided in \cref{an:spectral_entropy}, where we also further analyze their relationship to the informational content of the input configuration.

We emphasize that the cpVAE operates entirely unsupervised and does not rely on any prior knowledge of the system's Hamiltonian, symmetries, or phase boundaries. Despite this, it captures nontrivial structure in the data, organizing the snapshots into a meaningful latent geometry that, surprisingly, qualitatively matches the underlying phase diagram of the system. This demonstrates the power of probabilistic generative models to autonomously extract interpretable features from experimental quantum systems.

\section{Discussion}
In this work, we have presented a series of modifications on the VAE architecture essential to create interpretable representations of quantum data.  We have shown that two key aspects are crucial for this. First, as quantum data is intrinsically random, deterministic losses such as the mean square error will fail to capture the properties of most quantum systems. Instead, following common approaches in ML generative modeling, one rather needs to train the VAE to approximate the underlying probability distribution of the input data. Second, and closely related to the previous statement, the generative model must be able to generate data with same statistical properties as the input one. To this aim, we took inspiration from the recent advancements on neural quantum states (NQS), creating the VAE's decoder aligned with these architectures, ensuring that it can correctly reconstruct the quantum states under study.

By training different variants of VAEs in two paradigmatic spin models, we showed that failing to account for the intrinsic stochasticity of quantum data fundamentally limits the capacity of VAE to capture features beyond mean-field descriptors. In contrast, considering the aforementioned improvements enables the reconstruction of nontrivial correlation patterns and results in a latent space representation that matches the underlying structure of the quantum data. We then used such architecture to explore the phase space of snapshots obtain in an experimental setup of Rydberg atom arrays, demonstrating the ability of the model to uncover the structure of the underlying phase space and to distinguish between the different known quantum phases. Importantly, this is achieved in a fully unsupervised manner, with the model autonomously discovering latent factors that correspond to physically meaningful degrees of freedom.

Looking ahead, we note that in this study we restricted our analysis to spin models with a sign-positive ground state, so that measurement snapshots from a single computational basis sufficed. A natural next step is to lift this constraint and address systems with non-positive ground state, where faithful reconstruction will demand data drawn from several complementary measurement bases to capture the full quantum structure. Likewise, although our benchmarks focused on interacting two-level systems—$S=1/2$ spins and Rydberg atoms—the underlying methodology is readily transferrable to a broader class of platforms. With suitable decoder designs that reflect the statistics of the target system, the same prior-aligned VAE framework could be deployed to quantum circuits, $d$-level and/or multi-species.

The proposed framework can also be integrated with complementary interpretable techniques. For instance, equipping the encoder with correlator neural networks—architectures designed to capture designated local and non-local correlations in the data~\cite{miles2021correlator}—would allow the latent variables to be analysed in tandem with explicit correlation features, yielding a richer, more transparent description of the underlying physics. Furthermore, as demonstrated in \cref{sec:results_spins}, symbolic-regression methods offer a promising approach to extracting compact analytical expressions from the learned latents. The dimensionality and structure revealed by the cpVAE can in turn serve as informed priors for such regression algorithms, accelerating their convergence and enhancing the interpretability of the resulting formulae.

\section{Acknowledgments}

We sincerely thank Hannes Pichler, Sepehr Ebadi, Tout Wang, and Mikhail Lukin for providing the experimental data on the Rydberg atoms.

This research was funded in part by the Austrian Science Fund (FWF) [SFB BeyondC F7102, DOI: 10.55776/F71; WIT9503323, DOI: 10.55776/WIT9503323]. For open access purposes, the author has applied a CC BY public copyright license to any author accepted manuscript version arising from this submission. This work was also supported by the European Union (ERC Advanced Grant, QuantAI, No. 101055129). The views and opinions expressed in this article are however those of the author(s) only and do not necessarily reflect those of the European Union or the European Research Council - neither the European Union nor the granting authority can be held responsible for them.

\section{Code availability}
The code needed to reproduce the results presented in this work is available in Ref.~\cite{github}.
The datasets for the NNN-TFIM and the LR-TFIM were created by means of  NetKet~\cite{filippo2022netket, filippo2022codebase}, which is based on Jax~\cite{james2018jax} and MPI4Jax \cite{dion2021mpi4jax}. Netket has also been used for implementing the dense ARNN used as decoder of the cpVAE on the spin models. The implementation of the TC loss (\cref{eq:final_loss}) was done based on the implementation in~\cite{dubois2019disentangling}. The neural networks have been implemented using Flax~\cite{heek2024flax}.

\bibliography{bibiography}

\newpage

\onecolumngrid

\appendix

\section{Neural network architectures} \label{an:NN_archi}

In this section, we provide further details on the architecture of the neural network used for the encoder and decoder of the dVAE and the cpVAE through the different scenarios presented in the main text.

\subsection{Encoder and latent space}

For the VAEs trained on both spin models in ~\cref{sec:results_spins}, a convolutional neural network was used as the encoder. The details of its architecture are presented in \cref{tab:archi_convEncoder}. The same architecture is used for both dVAE and cpVAE. On the other hand, for the cpVAE trained on snapshots of Rydberg atom arrays (\cref{sec:results_Rybderg}), we used a transformer architecture for its encoder~\cite{vaswani2017attention}. Further details are presented in \cref{tab:archi_TransfEncoder}. 

In terms of the latent space, we consider for all cpVAE a latent dimension of $d_z=5$. We remind the reader that through training and minimization of \cref{eq:loss_VAE}, the model will autonomously noise out unnecessary neurons until it finds a minimal representation (i.e., 1 for the NNN-TFIM, 2 for LR-TFIM and 2 for the Rydberg atom dataset). We hence recommend training VAEs with large latent dimensions $d_z$, typically much larger than the expected degrees of freedom underlying the data. The model will then autonomously determine the optimal dimension.

\begin{table}[h]
    \centering
    \begin{tabular}{l|c}
    \hline
    \hline
       Layer type  & Output size \\
       \hline
       Input &  $B\times N \times 1$ \\
       Circular Conv 1D ($k_s=3$)  & $B\times N \times 32$ \\
       relu  & $B\times N \times 32$ \\
       Circular Conv 1D ($k_s=3$)  & $B\times N \times 32$ \\
       relu  & $B\times N \times 32$ \\
       Global Average Pooling  & $B\times 32$ \\
       \hline
       Mean predictions: &  \\
    \hspace{0.5cm} Dense & $B\times 64$ \\
    \hspace{0.5cm} relu & $B\times 64$ \\
    \hspace{0.5cm} Dense & $B\times 5$ \\
      Variance predictions: &  \\
    \hspace{0.5cm} Dense & $B\times 64$ \\
    \hspace{0.5cm} relu & $B\times 64$ \\
    \hspace{0.5cm} Dense & $B\times (d_z = 5)$ \\
    \hline
    \hline
       
    \end{tabular}
    \caption{\textbf{Convolutional encoder.} Details on the encoder architecture used for the NNN-TFIM and the LR-TFIM. $B$ is the batch size, $N$ the size of the spin configurations and $k_s$ the kernel size. The latent space dimension was chosen to be 5. The same architecture is used for the dVAE and the cpVAE.}
    \label{tab:archi_convEncoder}
\end{table}

\begin{table}[h]
    \centering
    \begin{tabular}{l|c}
    \hline
    \hline
       Layer type  & Output size \\
       \hline
       Input &  $B\times N \times 1$ \\
       Embedding  & $B\times N \times d_{\mathrm{model}}$ \\
       Positional Encoding ($\sin$) & $B\times N \times d_{\mathrm{model}}$ \\
       Transformer layer (3x):  &  \\
       \hspace{0.5cm} Normalization & $B\times N \times d_{\mathrm{model}}$ \\
       \hspace{0.5cm} Self-Attention ($N_{head}=d_{\mathrm{model}}/2)$ & $B\times N \times d_{\mathrm{model}}$ \\
        \hspace{0.5cm} Residual Connection & $B\times N \times d_{\mathrm{model}}$ \\
        \hspace{0.5cm} Dense & $B\times  4\cdot d_{\mathrm{model}} $ \\
        \hspace{0.5cm} relu & $B\times d_{\mathrm{model}} $ \\
        \hspace{0.5cm} Dense & $B\times N \times d_{\mathrm{model}}$ \\
        \hspace{0.5cm} Residual Connection & $B\times N \times d_{\mathrm{model}}$ \\
        \hspace{0.5cm} Normalization & $B\times N \times d_{\mathrm{model}}$ \\
        
       \hline
       Mean predictions: &  \\
    \hspace{0.5cm} Dense & $B\times d_{\mathrm{latent}}$ \\
      Variance predictions: &  \\
    \hspace{0.5cm} Dense & $B\times d_{\mathrm{latent}}$ \\
    \hline
    \hline
       
    \end{tabular}
    \caption{\textbf{Transformer encoder.} Details on the encoder architecture used for the cpVAE trained on snapshots of Rydberg atoms. $B$ in the batch size, $N$ is the number of atoms. A dimension of $d_{\mathrm{model}}=8$ was used. The latent space dimension was chosen to be 5.}
    \label{tab:archi_TransfEncoder}
\end{table}

\subsection{Autoregressive decoder} \label{an:autoregr_decoder}

As mentioned in the main text (see \cref{sec:methods}), autoregressive neural networks are used to model the conditional probabilities of each spin values
\begin{equation}
    p_{\theta}(x) = \prod_ip_{\theta}(x_i|x_0,x_1,...,x_{i-1}).
\end{equation}
Each conditional $p_{\theta}(x_i|x_0,...,x_{i-1})$ being normalized, this allow to sample spin configuration directly, without any Markov chain, typically needed for non-autoregressive models.

A common approach to render a neural network autoregressive is to impose a masking scheme that prevents access to future elements of the input sequence. For models trained on spin models (\cref{sec:results_spins}), we use as the decoder of the cpVAE a standard feedforward neural network (FFNN), with the masking being achieved by simply imposing triangular weights. For the numerical simulations, we use 3 hidden layers with 80 hidden neurons and \textit{selu} activation functions~\cite{klambauer2017self}. This design choice allowed us to employ the same neural network architecture for the decoder of the dVAE, omitting only the autoregressive masking used in the cpVAE. This ensures that comparisons between the two models remain as fair as possible.
 
Due to the higher complexity of the Rydberg atom datasets, we consider here a more powerful decoder based on the transformer architecture~\cite{vaswani2017attention} (see details in \cref{tab:archi_TransfDecoder}). The latent representation output by the encoder is fed into the transformer as context, following the typical context-based attention mechanism. This autoregressive approach requires an ordering of the 2D lattice atoms to be specified. We chose to use a snake ordering, which is commonly used for 2D systems~\cite{lange2024from, fitzek2024rydberggpt}.

\begin{table}[h]
    \centering
    \begin{tabular}{l|c}
    \hline
    \hline
       Layer type  & Output size \\
       \hline
       Input &  $B\times N \times 1$ \\
       Embedding  & $B\times N \times d_{\mathrm{model}}$ \\
       Positional Encoding ($\sin$) & $B\times N \times d_{\mathrm{model}}$ \\
       Shift & $B\times N \times d_{\mathrm{model}}$ \\
       Transformer layer (3x):  &  \\

       \hspace{0.5cm} Normalization & $B\times N \times d_{\mathrm{model}}$ \\
       \hspace{0.5cm} Causal Self-Attention $(N_{head}=d_{\mathrm{model}}/2)$ & $B\times N \times d_{\mathrm{model}}$ \\
       \hspace{0.5cm} Context Based Attention $(N_{head}=d_{\mathrm{model}}/2)$ & $B\times N \times d_{\mathrm{model}}$ \\
        \hspace{0.5cm} Residual Connection & $B\times N \times d_{\mathrm{model}}$ \\
        \hspace{0.5cm} Dense & $B\times  4\cdot d_{\mathrm{model}} $ \\
        \hspace{0.5cm} relu & $B\times  d_{\mathrm{model}} $ \\
        \hspace{0.5cm} Dense & $B\times N \times d_{\mathrm{model}}$ \\
        \hspace{0.5cm} Residual Connection & $B\times N \times d_{\mathrm{model}}$ \\
        \hspace{0.5cm} Normalization & $B\times N \times d_{\mathrm{model}}$ \\
        
       \hline
       Dense & $B\times N \times 2$ \\
       Log softmax & $B\times N \times 2$ \\

    \hline
    \hline
       
    \end{tabular}
    \caption{\textbf{Transformer decoder.} Details on the decoder architecture used for cpVAE trained on snapshots of the Rydberg atoms. $B$ is the batch size and $N$ the number of atoms. For the model presented in the main text, we fixed $d_{\mathrm{model}}=8$ and the latent space dimension to $d_{\mathrm{latent}}=5$. Not shown here, the context vector used in the context-based attention are the latent variables $z_i$, after a linear layer expanding their dimension to $B\times d_{\mathrm{latent}} \times d_{\mathrm{model}}$ }
    \label{tab:archi_TransfDecoder}
\end{table}

\section{Training details} \label{an:training}

\subsection{Hyperparameters}

In this section, we present the hyperparameters used for the numerical simulations. In particular, we will present the details on the training sets, the strength of each term in the loss (see \cref{eq:final_loss}) and the optimizer used for the gradient descent.

For both spin models, datasets were generated by sampling $10^4$ spin configurations from the exact ground state wavefunctions, computed using the Lanczos algorithm. For the NNN-TFIM, the Hamiltonian parameters $(J_2,h)$ (see \cref{eq:H_NNNTFIM}) were varied over a regular grid of  $(21,41)$ values. The coupling $J_2$ was linearly spaced in the interval $[0,1]$, while the transverse field $h$ was linearly spaced in $[0,2]$. For the LR-TFIM (see \cref{eq:H_LRTFIM}), the exponent $\alpha$ and transverse field $h$ were varied over a $(20,20)$ grid. The values of $\alpha $ were linearly spaced in $[1,5]$, whereas $h$ was logarithmically spaced in the range $[1.6,10]$.

For the experimental system of Rydberg atoms, we had access to $250$ snapshots taken for each set of parameters $(R_b/a,\Delta/\Omega)$ . In particular, the parameters were
\begin{equation}
    R_b/a \in \{1.01, 1.05, 1.13, 1.23, 1.3 , 1.39, 1.46, 1.56, 1.65, 1.71, 1.81, 1.89, 1.97\}
\end{equation}
and 31 values of $\Delta/\Omega \in [-2.33,4.65]$, linearly spaced.

The values used for the hyperparameters in the TC regularization (see \cref{eq:final_loss}) are presented in \cref{tab:TC_hyperparams_values}. A linear schedule was used for the weight on the component-wise KL, from $\gamma_{min}$ to $\gamma_{max}$.

\begin{table}[h]
    \centering
    \begin{tabular}{l|c|c|c|c}
    \hline
    \hline
       Spin model  & $\alpha$ & $\beta$ & $\gamma_{min}$ & $\gamma_{max}$ \\
       \hline
       NNN-TFIM & 0.1 & 30 & 0.1 & 0.2 \\
       LR-TFIM  & 0.1 & 0.5 & 0.5 & 10 \\
       Rydberg  & 0.001 & 10 & 1 & 1 \\
       
    \hline
    \hline
       
    \end{tabular}
    \caption{\textbf{TC regularization hyperparameters} Values of the hyperparameters used in \cref{eq:final_loss} for training the cpVAE on the two spin models and on the Rydberg atom arrays snapshots.}
    \label{tab:TC_hyperparams_values}
\end{table}

We then optimized the neural network with the \textit{adabelief} optimizer~\cite{zhuang2020adabelief}, with default parameters and a learning rate of 0.001. The training stops when the reconstruction loss and the variance of the latent variables converge.

\subsection{Training}

\begin{figure}[h]
    \centering
    \includegraphics[width=1.0\linewidth]{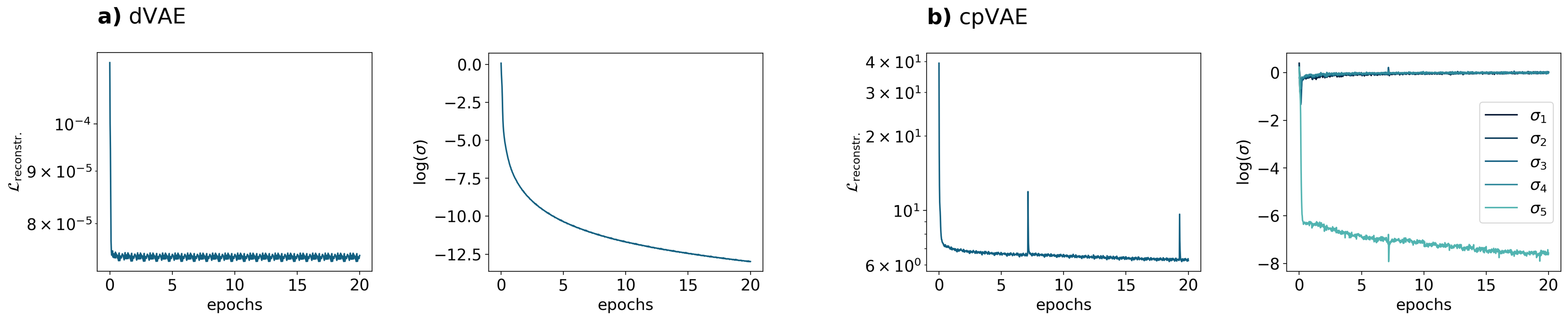}
    \caption{\textbf{Training: NNN-TFIM}. Training of a) the dVAE and b) the cpVAE seen through the reconstruction loss $\mathcal{L}_{\mathrm{reconstr.}}$ and the $\log$ of the variances $\sigma_i$ output by the encoder. For the dVAE the reconstruction loss is the mean square error (MSE) and for the cpVAE it is the first term in \cref{eq:final_loss}.}
    \label{fig:NNNTFIM_training}
\end{figure}

\begin{figure}[h]
    \centering
    \includegraphics[width=1.0\linewidth]{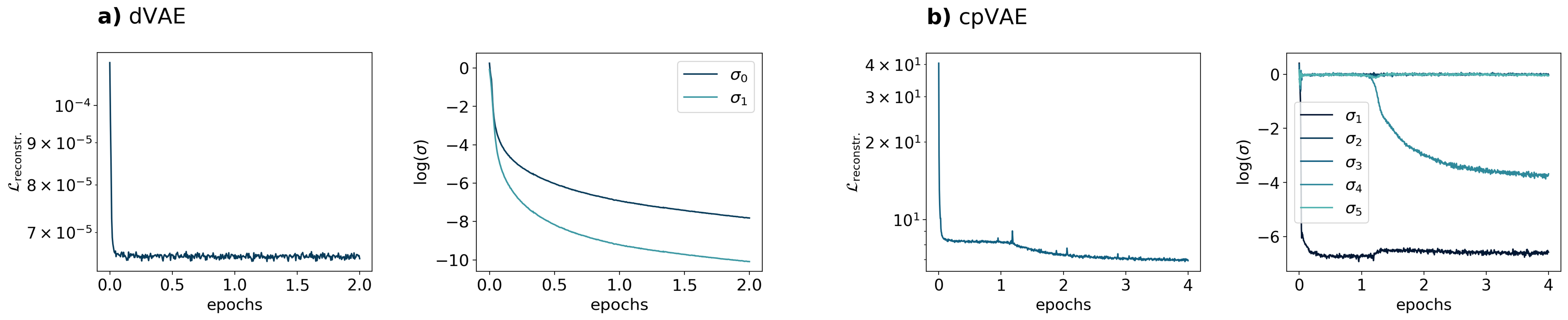}
    \caption{\textbf{Training: LR-TFIM}. Training of a) the dVAE and b) the cpVAE seen through the reconstruction loss $\mathcal{L}_{\mathrm{reconstr.}}$ and the $\log$ of the variances $\sigma_i$ output by the encoder. For the dVAE the reconstruction loss is the mean square error (MSE) and for the cpVAE it is the first term in \cref{eq:final_loss}.}
    \label{fig:LRTFIM_training}
\end{figure}

In this section, we provide additional details on the training of the dVAE and the cpVAE. The training dynamics of both VAEs on the NNN-TFIM and LR-TFIM are presented in \cref{fig:NNNTFIM_training} and \cref{fig:LRTFIM_training}, respectively. We show the evolution of the reconstruction loss $\mathcal{L}_{\mathrm{reconstr.}}$ and the logarithm of the latent variances $\log \sigma_i$ output by the encoder.

For the dVAE, the reconstruction loss corresponds to the mean squared error (MSE). Across both spin models, this loss decreases rapidly in the early training stages and subsequently plateaus. This typically indicates that the model has insufficient capacity or expressivity to fully learn the data distribution \cite{smith2018disciplinedapproachneuralnetwork}. In our case, as shown in the main text, this model was not able to approximate the quantum probability distribution and capture the correlations. 

In order to ensure a fair comparison between the dVAE and cpVAE, the dVAE was trained without KL regularization. Its latent dimension was fixed to match the expected number of relevant latent factors, as found by the cpVAE and previously known for the presented benchmark cases. Under these conditions, we observe that the variances $\log \sigma_i$ decrease rapidly for both spin models.

When training without KL regularization and restricting the latent dimension, our goal is to design the fairest possible comparison between the dVAE and the proposed cpVAE. As discussed in the main text, the main purpose of the KL divergence term is to minimize the number of active neurons in the latent space to create a minimal representation of the input data. This is the main challenge in VAE training and typically depends on good hyperparameterisation of the weights in~\cref{eq:final_loss}.

With the proposed approach, this objective (learning a minimal latent representation) is achieved directly by constraining the latent dimensionality of the dVAE. Introducing the KL term would impose additional constraints on the latent space of the dVAE and could reduce the expressiveness of the latent representation. Furthermore, tuning the KL weight introduces additional hyperparameters, which complicates the comparison and risks introducing bias in favor of one model. By avoiding the KL term, we consider the 'best possible' scenario for the dVAE, which still fails to match the representation power of the cpVAE.

Meanwhile, the KL term, and more importantly, its TC decomposition (see~\cref{eq:TC}), helps the model find more interpretable representations by promoting uncorrelated latent neurons, a feature commonly known as disentanglement. However, it should be noted that the aim here is not to compare the structure of the latent representations, but rather to demonstrate that the dVAE cannot learn parts of the phase space due to its deterministic nature, in contrast to the cpVAE.

For the cpVAE, the reconstruction loss corresponds to the first term in \cref{eq:final_loss}. During training, this loss exhibits a rapid initial decrease followed by a more gradual decline, accompanied by small fluctuations characteristic of stochastic mini-batch optimization. The evolution of the logarithmic variances $\log \sigma_i$ for the five latent neurons is also displayed. In the case of the NNN-TFIM, the model selectively activates a single latent neuron, indicated by a decrease in its $\log \sigma_i$, while effectively suppressing the remaining neurons by driving their variances close to zero. This behavior reflects a sparsity in the learned latent representation, consistent with the existence of a dominant underlying feature in the data. For the LR-TFIM, the model retains two active latent neurons. Notably, the activation of the second latent neuron occurs after approximately one epoch of training, enabling the model to further reduce the reconstruction loss. This dynamic reflects the model's ability to progressively allocate capacity as needed to capture increasingly complex structure in the data.

\begin{figure}[h]
    \centering
    \includegraphics[width=0.55\linewidth]{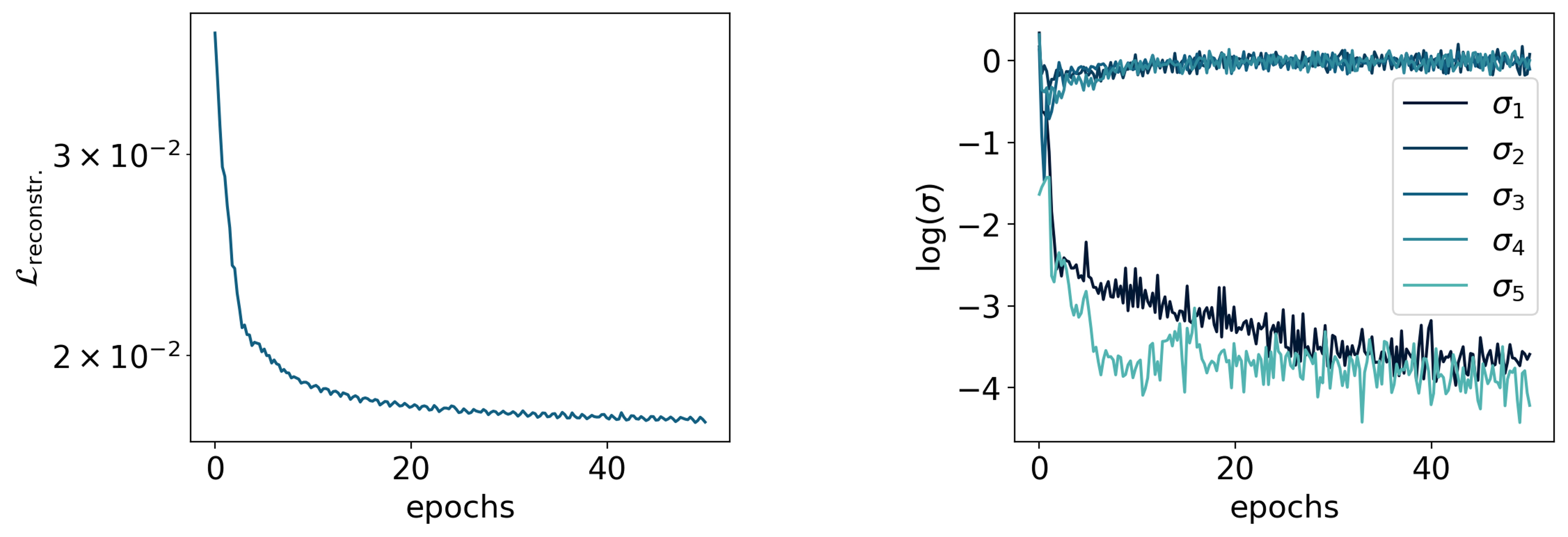}
    \caption{\textbf{Training: Rydberg atoms}. Training of the cpVAE seen through the reconstruction loss $\mathcal{L}_{\mathrm{reconstr.}}$ (left) and the $\log$ of the variances $\sigma_i$ output by the encoder (right).}
    \label{fig:Rydberg_training}
\end{figure}

Finally, we present the training dynamics of the cpVAE on experimental snapshots of Rydberg atoms in \cref{fig:Rydberg_training}. The reconstruction loss decreases smoothly over time, and the model rapidly converges to a representation involving two active latent neurons, as evidenced by the corresponding drop in $\log \sigma_{1,5}$. The remaining neurons are effectively noised out by pushing their variances near zero, again indicating a sparse and interpretable latent encoding.

\section{NNN-TFIM: Generalization abilities} \label{an:NNNTFIM_generalize}

\begin{figure}[h]
    \centering
    \includegraphics[width=0.6\linewidth]{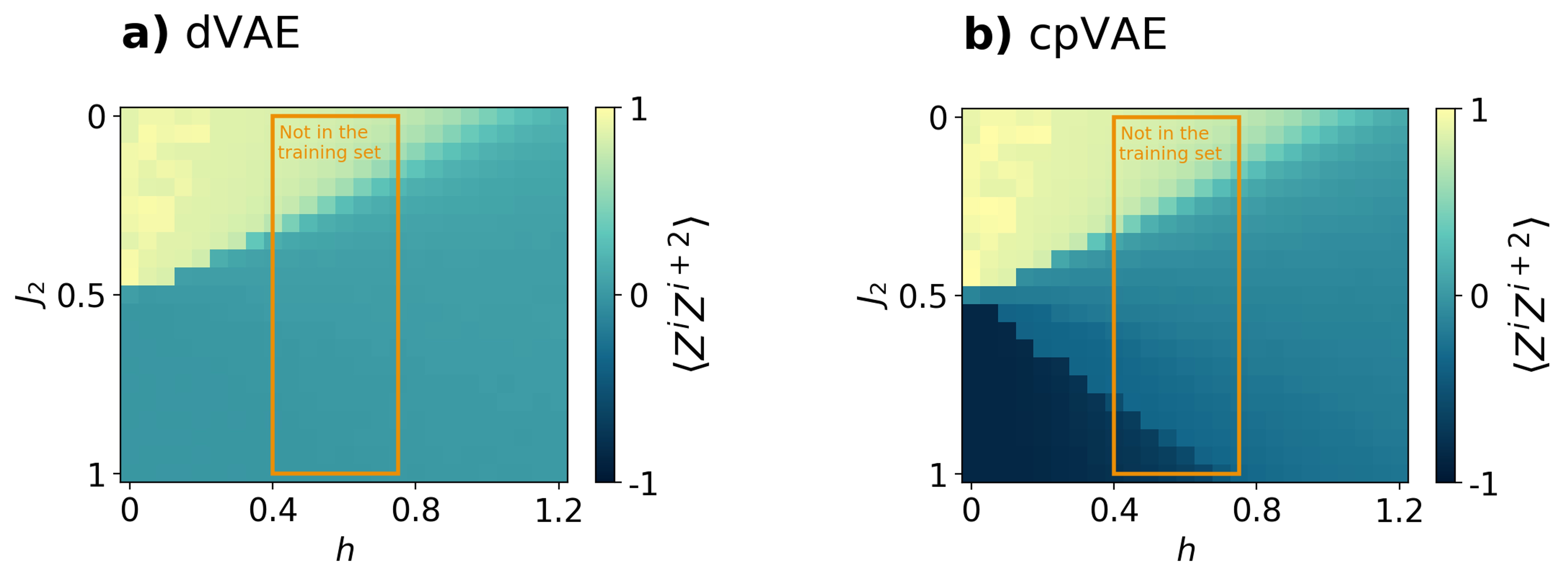}
    \caption{\textbf{Generalization abilities: NNN-TFIM}. Next-nearest-neighbor correlator of the spin configuration generated by a) the dVAE and b) the cpVAE. The training of both VAE were performed without spin configurations coming from the ground state of Hamiltonians having $0.4<h<0.75$, region inside the orange rectangle. Then they are tested on the entire space.}
    \label{fig:NNNTFIM_reconstr_corr_hole}
\end{figure}

The generalization capabilities of the dVAE and cpVAE were also investigated on the NNN-TFIM using a restricted training set. Specifically, both models were trained without any spin configuration coming from the region $0.4<h<0.75$ of the parameter space. We then evaluated the generation performance of both models across the entire phase space, including the excluded region. The reconstructed values of the nearest-neighbor correlator are presented in \cref{fig:NNNTFIM_reconstr_corr_hole}a and b for the dVAE and cpVAE, respectively. Notably, the quality of the reconstructed correlators in the previously unseen region remains consistent with that of the training regions. This indicates that both architectures exhibit robust generalization to regions of the phase space that were not included in the training set.

\section{LR-TFIM: Sampling from the latent space} \label{an:LRTFIM_sampling}

\begin{figure}[h]
    \centering
    \includegraphics[width=0.9\linewidth]{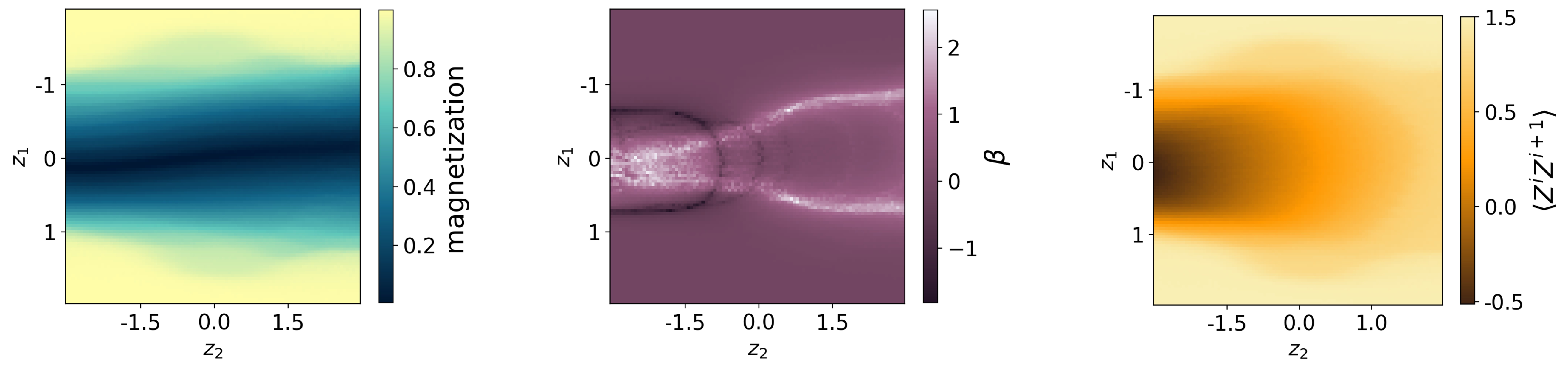}
    \caption{\textbf{Generation: LR-TFIM}. To further interpret the structure learned by the cpVAE, a scan is performed over the latent variables $z_{1,2}$. Then, for each values of the latter, samples are generated using the decoder of the cpVAE. Left is the magnetization of these samples, middle is the $\beta$ exponent and right is the correlators $\langle Z^iZ^{i+1}\rangle$.}
    \label{fig:LRTFIM_sampling_from_latent_space}
\end{figure}

To further interpret the structure learned by the cpVAE, we leverage its generative nature to probe the contents of the latent space. A linear sweep is performed over each latent dimension independently, and for every point along the scan, 1000 spin configurations are sampled from the generative decoder. For each set of generated configurations, we compute the magnetization, the correlation exponent $\beta$ and the nearest neighbor correlator $\langle Z^iZ^{i+1}\rangle$. The results are presented in \cref{fig:LRTFIM_sampling_from_latent_space}.
This generative probing again highlights that the two latent dimensions encode distinct physical features: the first latent neuron predominantly controls the magnetization of the output configurations, while the second  mainly governs their correlation. Notably, for certain latent values, the model generates spin configurations with an apparent $\beta < 0$. These correspond to atypical states that exhibit very weak or inverted correlation patterns, in which nearest-neighbor correlations vanish and increase slightly with distance. In these cases, the power-law fit used to estimate $\beta$ becomes unreliable.

\section{Symbolic regression methods to extract analytical expressions} \label{an:SR_LRTFIM}

As mentioned in the main text, for the LR-TFIM, the second active latent neuron of the cpVAE highlighted the presence of an intermediate phase at low values of $\alpha$ and $h$. To further investigate the physical nature of this emergent region, we applied symbolic methods. In particular, we used symbolic regression (SR), implemented via the open-source PySR package~\cite{crammer2023interpretable}, to extract closed-form analytical relations that describe this additional phase in the latent representation.

PySR is based on a multi-population evolutionary algorithm that evolves sets of candidate expressions through repeated mutation, crossover and simplification. The algorithm refines these expressions iteratively using a loop comprising the following three steps: (i) new candidates are evolved via symbolic mutations; (ii) expressions are simplified using algebraic rules to avoid redundant complexity; and (iii) scalar constants are optimized using gradient-based solvers. This loop enables PySR to discover compact analytical relations containing unknown real parameters, which is crucial for scientific applications.

We design the SR to find a symbolic expression $f(\mathbf{x})$ that distinguishes the spin configurations $\mathbf{x}$ that come from the additional phase (inducing $\mu_2>0.35$) from those outside. The best expressions we recovered are dominated by two-body correlator terms with alternating signs. For example, one extracted expression (truncated) was
\begin{equation*}
    f(\mathbf{x}) = -x_0 x_1-2x_0x_1-x_0x_{12}-x_0x_{14}-x_0x_{16}-4x_0x_{17}+2x_0x_{12}+x_0x_{19}+4.5x_0x_2-3.2x_0x_3+x_0x_8+...
\end{equation*}
Although studying these expressions led us to the correct structure factor parameter, these fits were neither robust nor easily interpretable, hence we have omitted them from the main results (see~\cref{eq:struc_factor}). Nevertheless, as mentioned in the conclusion, we regard the combination of learned latent representations with symbolic regression as a promising direction, although we leave its proper integration within the VAE pipeline for future work.

\section{Study of the latent neuron variances for spin models benchmarks} \label{an:spectral_entropy}

In the main text, we discuss the behavior of the latent variances $\sigma_i$ output by the cpVAE encoder when trained on Rydberg atom snapshots (see \cref{sec:results_Rybderg}). Here we present an extended analysis of the variances output by the encoder of the cpVAE trained on the spin models (see \cref{sec:results_spins}).

As commented in the main text, the variances output by the encoder reflect the permissible level of noise in the sampling of the latent variables $z_i$ such that the decoder can still accurately reconstruct the input. Empirically, we observe that these variances tend to correlate with the degree of apparent randomness in the input data. To quantify this relationship, we compare the learned variances with the spectral entropy of the input spin configurations. Spectral entropy is a scalar metric that captures the randomness or complexity of a signal, and is particularly well-suited for binary data such as spin configurations. The core idea is that structured signals concentrate their spectral power in specific frequency components, whereas random signals exhibit a more uniform distribution across the frequency domain. Consequently, the spectral entropy serves as a compact and informative measure of the underlying order or disorder in the input data.

From a practical perspective, the spectral entropy compresses the normalized power spectral density of a signal into a single scalar value, and is computed using the Shannon entropy of its frequency distribution. In this framework, lower spectral entropy corresponds to more structured, periodic, input configurations, while higher values indicate increased stochasticity and disorder. This metric has been employed across various domains, including molecular component identification~\cite{li2021spectral}, the characterization of chaotic dynamics~\cite{xiong2021spectral} and classification tasks~\cite{swetapadma2023novel}, demonstrating its broad applicability for quantifying structure and randomness in complex data.

To compute such metric, we begin by applying a fast Fourier transform (FFT) to the spin configuration. The power spectrum is then obtained by taking the squared magnitude of the FFT coefficients. This spectrum is normalized such that it can be interpreted as a probability distribution $I_p$, from which the Shannon entropy $S$ is calculated as
\begin{equation}
    S = - \sum_p I_p \log I_p.
\end{equation}

\begin{figure}[h]
    \centering
    \includegraphics[width=0.9\linewidth]{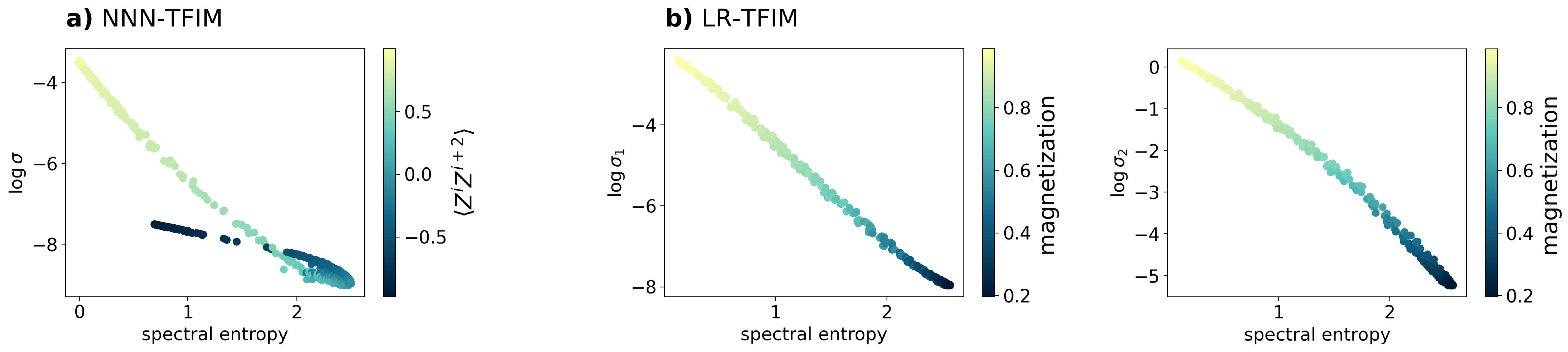}
    \caption{\textbf{Behavior of the latent variances $\sigma$ for the NNN-TFIM and the LR-TFIM}. We show here the output $\sigma_i$ of the cpVAE encoder as a function of the spectral entropy of the input spin configurations. \textbf{a)} NNN-TFIM, where a single latent neuron remains active after training. \textbf{b)} LR-TFIM, for which two latent neurons remained active.}
    \label{fig:logvar_vs_spectral_entropy}
\end{figure}

The values of $\sigma_i$ output by the encoder of the cpVAE are shown in \cref{fig:logvar_vs_spectral_entropy} as a function of the spectral entropy of the input spin configuration. Overall, a linear dependence between their logarithm $\log\sigma_i$ and the spectral entropy is observed. This trend can be understood through the competing objectives of the VAE’s loss function~\cref{eq:loss_VAE}: the reconstruction term $\mathcal{L}_{\mathrm{reconstruction}}$ versus the Kullback–Leibler regularization term $\mathcal{L}_{\mathrm{KL}}$. When the input exhibits a high degree of regularity and is easier to predict, the encoder can output smaller $|\log\sigma_i|$ values, thereby reducing the KL contribution to the total loss. Conversely, inputs with higher spectral entropy—indicative of greater randomness—necessitate larger encoder variances to preserve sufficient flexibility in the latent representation for accurate reconstruction.

Notably, for the NNN-TFIM, this relationship exhibits a piecewise linear or "two-regime" behavior. Specifically, spin configurations from the antiferromagnetic phase require systematically higher latent variances compared to those from the ferromagnetic phase, even when their spectral entropy values are comparable. This observation suggests that the antiferromagnetic phase encodes richer, more intricate structure, requiring the model to allocate greater representational capacity to faithfully encode these configurations.    

\end{document}